\documentclass[reprint,superscriptaddress]{revtex4-1}
\usepackage[utf8]{inputenc}

\usepackage{amsmath}
\usepackage{bm}
\usepackage{graphicx}
\usepackage{sidecap}
\usepackage{xcolor}
\usepackage{hyperref}
\usepackage{ulem}
\usepackage{enumitem}
\usepackage{comment}
\usepackage{tikz}

\DeclareMathOperator{\sech}{sech}
\newcommand{\p}{\partial}
\newcommand{\half}{\frac{1}{2}}

\newcommand{\ex}{\bm{\hat{e}}_1}
\newcommand{\ey}{\bm{\hat{e}}_2}
\newcommand{\ez}{\bm{\hat{e}}_3}

\newcommand{\emu}{\bm{\hat{e}}_\mu}
\newcommand{\emn}{\epsilon_{\mu\nu}}
\newcommand{\dx}{\mathrm{d}x}
\newcommand{\dy}{\mathrm{d}y}

\newcommand{\nagn}{\bm{n}}
\newcommand{\heff}{\bm{f}}
\newcommand{\nDW}{\nagn_{\rm DW}}

\newcommand{\dm}{\lambda}

\newcommand{\Energy}{E}
\newcommand{\Potential}{V}
\newcommand{\vel}{v}
\newcommand{\width}{w}

\begin{document}

\title{Vortex propagation and phase transitions in a chiral antiferromagnetic nanostripe}
\author{Riccardo Tomasello}
\affiliation{Institute of Applied and Computational Mathematics, FORTH, Heraklion, Crete, Greece}
\author{Stavros Komineas}
\affiliation{Institute of Applied and Computational Mathematics, FORTH, Heraklion, Crete, Greece}
\affiliation{Department of Mathematics and Applied Mathematics, University of Crete, 70013 Heraklion, Crete, Greece}
\date{\today}

\begin{abstract}
We study a vortex in a nanostripe of an antiferromagnet with easy-plane anisotropy and interfacial Dzyloshinskii-Moriya interaction.
The vortex has hybrid chirality being N\'eel close to its center and Bloch away from it.
Propagating vortices can acquire velocities up to a maximum value that is lower than the spin wave velocity.
When the vortex is forced to exceed the maximum velocity, phase transitions occur to a nonflat spiral, vortex chain, and flat spiral, successively.
The vortex chain is a topological configuration stabilised in the stripe geometry.
Theoretical arguments lead to the general result that the velocity of localized excitations in chiral magnets cannot reach the spin wave velocity.
\end{abstract}

\maketitle

\section{Introduction}
\label{sec:intro}

A wide range of materials present antiferromagnetic order, where neighboring magnetic moments are coupled via a strong exchange interaction and are aligned antiparallel.
Antiferromagnets (AFMs) exhibit features, such as low magnetic susceptibility, robustness against external fields and lack of stray fields, that are favorable for the building blocks of spintronic devices  \cite{Jungwirth2016, Baltz2018}.
They receive renewed interest because current techniques allow for the antiferromagnetic order to be manipulated by spin-currents and to be observed despite the lack of net magnetization \cite{Wadley2016, Grzybowski2017, Moriyama2018, Bodnar2019, Baldrati2019, Shi2020}.
This opens the way for a number of potential applications including storage with picosecond switching \cite{Cheng2015, Roy2016, Lopez-Dominguez2019}, THz oscillators \cite{Cheng2016, Khymyn2017, Puliafito2019}, racetrack memory based on magnetic solitons such as domain walls (DWs) \cite{Gomonay2016, Shiino2016, Sanchez-Tejerina2020} or skyrmions \cite{Zhang2016, Barker2016, Gomonay2018, Salimath2020}, which can achieve velocities larger than 1 km/s \cite{Gomonay2016, Shiino2016, Salimath2020}.

Some AFM materials such as $\alpha-{\rm Fe}_2{\rm O}_3$, ${\rm Ba}_2{\rm CuFe}_2{\rm O}_7$, are characterized by easy-plane anisotropy which supports the formation of vortices.
They have been discussed theoretically in infinite films \cite{IvanovSheka_PRL1994, Pereira1995, Ivanov1996, Bogdanov1998, Komineas1998}
and observed experimentally by imprinting techniques \cite{Wu2011, Chmiel2018}.
Despite of that, they have received much less attention than DWs or skyrmions or also than their ferromagnetic counterparts \cite{1989_PRB_GouveaWysinBishopMertens, PapanicolaouSpathis_NL1999, 2000_Science_Shinjo, 2002_Science_Wachowiak, WaeyenbergePuzic_Nat2006, Yamada2007, Pribiag2007, Komineas2007}. 
%%Maybe V. I. Pokrovskii and D. V. Khveshchenko, Sov. J. Low Temp. Phys. 14, 213 (1988)

An extensive experimental investigation of an easy-plane AFM with the Dzyaloshinskii-Moriya interaction (DMI) established spiral antiferromagnetic order \cite{2011_PRB_MuhlbauerZheludev, 2012_PRB_MuhlbauerZheludev}
and a subsequent theoretical analysis has shown the existence of two spiral phases \cite{2002_PRB_ChovanPapanicolaou, 2005_springer_ChovanPapanicolaou}.
For weak DMI, the N\'eel is the ground state, but for stronger DMI the system enters a spiral phase where all N\'eel vector components vary in space (nonflat spiral).
Only for strong enough DMI the N\'eel vector lies in a plane and rotates in space thus giving a flat spiral.
The nonflat spiral gives an intermediate phase that is not there in the case of an easy-axis magnet \cite{BogdanovHubert_JMMM1999}.

We study theoretically vortices in easy-plane AFMs with an interfacial DMI.
We consider a stripe geometry as this is the most suitable for applications involving shifting of magnetic information, while it will also give rise to interesting effects on the magnetic structure.
We calculate the magnetic ground state and demonstrate that this induces a vortex with a mixed chirality, i.e., N\'eel-type near the vortex core and Bloch-type away from it.
This unusual type of vortex will be referred to as a {\it hybrid} vortex.

We subsequently study propagating vortices.
We show that a propagating vortex shrinks along the direction of propagation, similarly to AFM DWs \cite{Shiino2016}, while it elongates along the perpendicular direction, similarly to AFM skyrmions \cite{Salimath2020, KomineasPapanicolaou_SciPost2020}.
The vortex can acquire a maximum velocity beyond which it becomes unstable to periodic configurations, thus giving rise successively to a nonflat spiral, a vortex chain and a flat spiral.
The spirals are extensions of states known within the one dimensional model, but the vortex chain is a feature of the stripe geometry.
A theoretical explanation for the dynamical behavior is obtained and it leads to the general result that the velocity of localized excitations in chiral magnets cannot reach the spin wave velocity.
Our results provide an understanding of the statics and dynamics of vortices in chiral AFMs and could be useful for the design of antiferromagnetic devices based on magnetic solitons.

%\textcolor{red}{
%The paper is organised as following.
%In Sec.~\ref{sec:formulation}, we ...
%}

%\todo[inline, color=green!40]{We can use notes.}

\begin{figure*}[t]
    \centering
    \includegraphics[width=17.5cm]{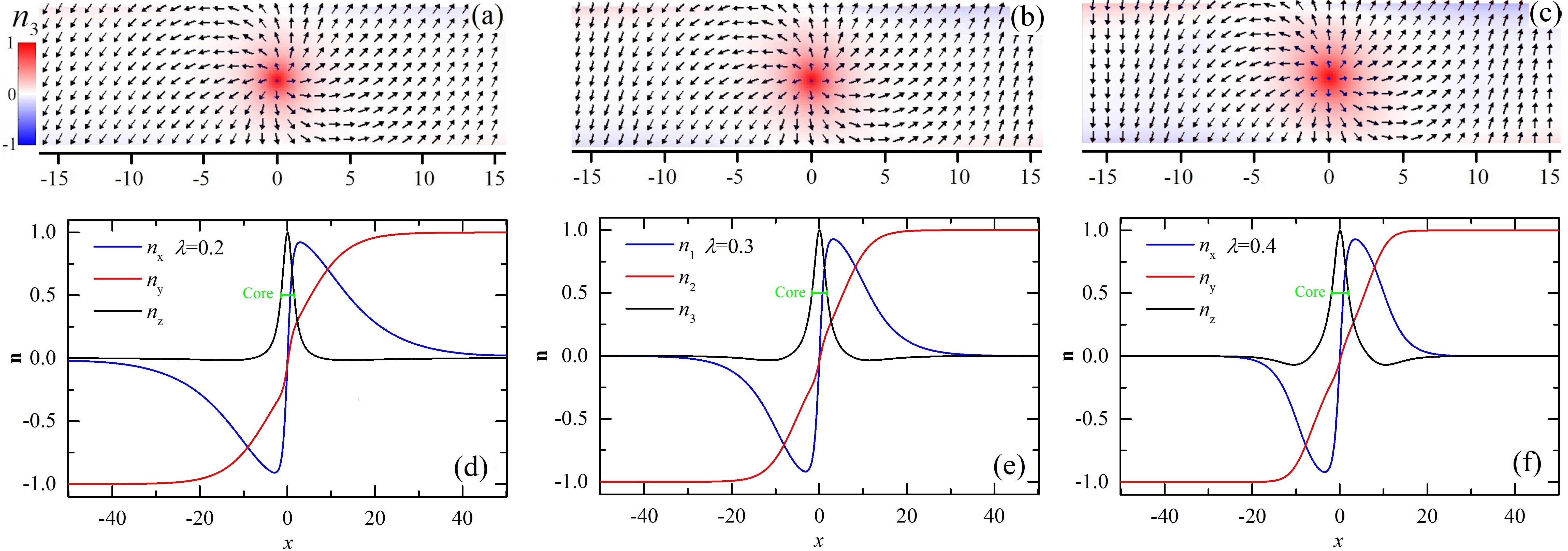}
    \caption{Static hybrid vortex in a stripe with width $\width=10$.
    The length of the numerical mesh along $x$ is $L=100$.
    (a), (b) ,(c) Vector plot of the static hybrid vortex for three values of the DMI parameter.
    Vectors show the projection of the N\'eel vector on the plane, $(n_1, n_2)$ while the component $n_3$ is shown by a color code.
    (d), (e) ,(f) The components of $\nagn$ along the line in the center of the stripe ($y=0$) for the configurations shown in (a), (b), (c) respectively.
    The vortex core width is shown by a green solid line.
    }
    \label{fig:vortexProfiles}
\end{figure*}

\section{The model and ground states}
\label{sec:formulation}

%\paragraph*{Continuum model.}
We consider an antiferromagnetic nanostripe with exchange, interfacial DMI and easy-plane anisotropy.
A continuum model is obtained for the normalized N\'eel vector $\nagn = (n_1,n_2,n_3)$ \cite{BaryakhtarIvanov_SJLTP1979,KomineasPapanicolaou_NL1998} with the potential energy
\begin{align} \label{eq:potential_iDMI}
 \Potential = \int & \left[ \half (\p_\mu\nagn)\cdot(\p_\mu\nagn) \right. \\
 & \left. - \dm \emn \emu\cdot(\p_\nu\nagn\times\nagn) 
 + \frac{1}{2} n_3^2 \right] \dx\dy, \notag
\end{align}
where $\mu,\nu$ take the values 1,2, $\emn$ is the totally antisymmetric tensor, $\emu$ denote the unit vectors in the respective directions, and $\dm$ is a scaled DMI parameter.
The equation of motion is
\begin{align}  \label{eq:sigmaModel}
& \nagn\times ( \ddot{\nagn} - \heff ) = 0,  \\
& \heff = \Delta\nagn + 2\dm \emn \emu\times\p_\nu\nagn - n_3\ez. \notag
\end{align}
%The magnetization is given in terms of the N\'eel vector
%\begin{equation}
%    \magn = \frac{\epsilon}{2\sqrt{2}} \nagn\times\dot{\nagn}
%\end{equation}
%where $\epsilon$ is a small parameter that can be taken to be the spin lattice spacing.

Let us review the results for a one dimensional (1D) model with the energy \eqref{eq:potential_iDMI} and $\nagn=\nagn(x)$.
Phase transitions occur at the two critical values of the parameter \cite{2002_PRB_ChovanPapanicolaou}
\begin{equation} \label{eq:dm1_dm2}
\dm_{NF} = \frac{1}{2},\qquad
\dm_F \approx 0.705. 
\end{equation}
We give schematically the three regimes separated by the critical values of $\dm$.

\begin{center}
\begin{tikzpicture}
  \draw[thick,->] (0,0) -- node[above]{N\'eel} ++(2,0) -- node[below]{$\dm_{NF}$} ++(0.5,0) -- node[above]{nonflat spiral} ++(2.5,0) -- node[below]{$\dm_F$} ++(0.5,0) -- node[above]{flat spiral} ++(2,0) --
  node[below]{$\dm$} ++(1,0);
  \draw[fill] (2.1,0) circle [radius=0.05];
  \draw[fill] (5.1,0) circle [radius=0.05];
\end{tikzpicture}
\end{center}

For weak DMI, $\dm < \dm_{NF}$, the N\'eel state is the ground state (see Ref.~\cite{Tomasello2020} for a related model).
%%Is the above Ref relevant?
The N\'eel vector is lying in the easy-plane and, for definiteness, we will assume that this is $\nagn=\ey$.
Increasing $\dm$, we enter an intermediate phase in the form of a nonflat spiral at $\dm=\dm_{NF}$.
The spiral presents a continuous rotation of the projection of $\nagn$ on the $(13)$ plane as we move along the $x$ axis and, at the same time, the component $n_2$ oscillates around a nonzero value.
The period of the spiral tends to infinity for $\dm\to\dm_{NF}$ while the component $n_2\to1$ in the same limit.
As $\dm$ increases, $n_2$ decreases and it vanishes at $\dm=\dm_F$ where a flat spiral is obtained with $\nagn$ lying fully and rotating on the $(13)$ plane.
For $\dm > \dm_F$, the flat spiral remains the ground state and the period of the spiral decreases with increasing $\dm$ \cite{2002_PRB_ChovanPapanicolaou, Tomasello2020}.

\section{Vortex in a stripe}
\label{sec:stripe}

Let us now assume a stripe geometry. This extends to infinity along the $x$ axis and it has a width $\width$ in the $y$ direction, $-\width/2 \leq y \leq \width/2$.
We focus on the regime $\dm < \dm_{NF}$ where we expect a N\'eel state.
Any solution of Eq.~\eqref{eq:sigmaModel} should satisfy the natural boundary condition
\begin{equation} \label{eq:bc_1D}
    \p_y\nagn + \dm \ex\times\nagn = 0,\quad
    y=\pm \frac{\width}{2}.
\end{equation}

In the finite interval $-\width/2 < y < \width/2$, two degenerate nontrivial ground states with negative energy can be found, as shown in Appendix~\ref{sec:groundState}.
We denote these $\nagn=\nagn_\pm$ and $\nagn$ is primarily aligned along $\pm\ey$.
In the case of the stripe, we extend the previous 1D configuration in the $x$ direction and we have two degenerate ground states where $\nagn$ does not depend on $x$, that is $\nagn(x,y)=\nagn_\pm(y)$.
This is a quasi-uniform state where the N\'eel vector points primarily along $\pm\ey$ and it tilts out of the plane, in $\ez$, in the regions close to the boundaries $y=\pm \width/2$.
One can say that the boundary condition \eqref{eq:bc_1D} makes $\ey$ an energetically favorable axis.

We simulate the system numerically on a stripe domain with a long $x$ dimension, that typically contains $1000$ grid points with lattice spacing 0.1, giving a physical dimension $100$.
We vary the width of the stripe.
We impose Neumann boundary conditions at the ends of the numerical mesh in the $x$ direction.
In the $y$ direction, we use open boundary conditions at $y=\pm\width/2$. (In Appendix~\ref{sec:groundState}, it is shown that these give the same results as the natural boundary conditions).
A relaxation algorithm indeed converges to a quasi-uniform state of the form $\nagn=\nagn_\pm(y)$, that does not depend on $x$.

Vortices should be excited states on the quasi-uniform state in the regime $\dm < \dm_{NF}$.
Due to the form of the DMI, a vortex solution of model \eqref{eq:sigmaModel} is expected to be of N\'eel type in an infinite film.
The form of the ground state forces us to assume in-plane domains oriented primarily along the $\pm\ey$ on the left and right side of the stripe, respectively, i.e., $\nagn(x\to\pm\infty,y) = \nagn_\pm(y)$, separated by an out-of-plane domain wall in the center of the stripe.
This ansatz is used as an initial condition in our numerical relaxation method.
We run simulations for different widths $\width$ and parameter values $\dm$.
A vortex is obtained as an equilibrium state for stripes with width larger than a critical width that depends on $\dm$.
For $\width>4$, we obtain a vortex for all values of $\dm$.

Figure~\ref{fig:vortexProfiles} shows the results of simulations on a $1000\times100$ grid with lattice spacing 0.1, giving physical dimensions $100\times10$.
%We will assume a vortex with polarity up, i.e., $\nagn=\ez$ in the center of the vortex, in all our calculations.
In Fig.~\ref{fig:vortexProfiles}, the entries (a), (b), (c) show vector plots of a static vortex in a stripe with $\width=10$ for three values of the DMI parameter $\dm=0.2, 0.3, 0.4$.
The vortex is N\'eel close to the vortex core and it gradually becomes Bloch as we go away from the core thus exhibiting a hybrid character.
Starting from the vortex core, the N\'eel vector goes towards the in-plane direction by rotating in the $(13)$ as well as in the $(23)$ plane.
This results in a vortex configuration which is between N\'eel and Bloch near the vortex core, similarly to what happens with Dzyaloshinskii DWs \cite{Thiaville2012} or skyrmions with intermediate chirality \cite{Buttner2018, Olleros-Rodriguez2020}.
%The simultaneous rotation occurring in the two planes can be ascribed to two effects:
The $(13)$ rotation is a consequence of the interfacial DMI, while the $(23)$ rotation is a consequence of the boundary conditions which force the N\'eel vector to be oriented primarily along $\ey$ in the far field.
As we move further from the vortex core, the magnetization becomes aligned with $\pm\ey$ in opposite directions on the left and right side of the stripe.

In Fig.~\ref{fig:vortexProfiles}, the entries (d), (e), (f) show the N\'eel vector profiles along the line in the center of the strip, $y=0$, corresponding to the vortices in entries (a), (b), (c), respectively.
Increasing the DMI parameter has two main effects: (i) an increase of the vortex core width $L_0$, as also noted in the Appendix of Ref.~\cite{2002_PRB_ChovanPapanicolaou}, and (ii) a faster rotation of the N\'eel vector towards $\ey$.

\begin{figure}
    \centering
    \includegraphics[width=8cm]{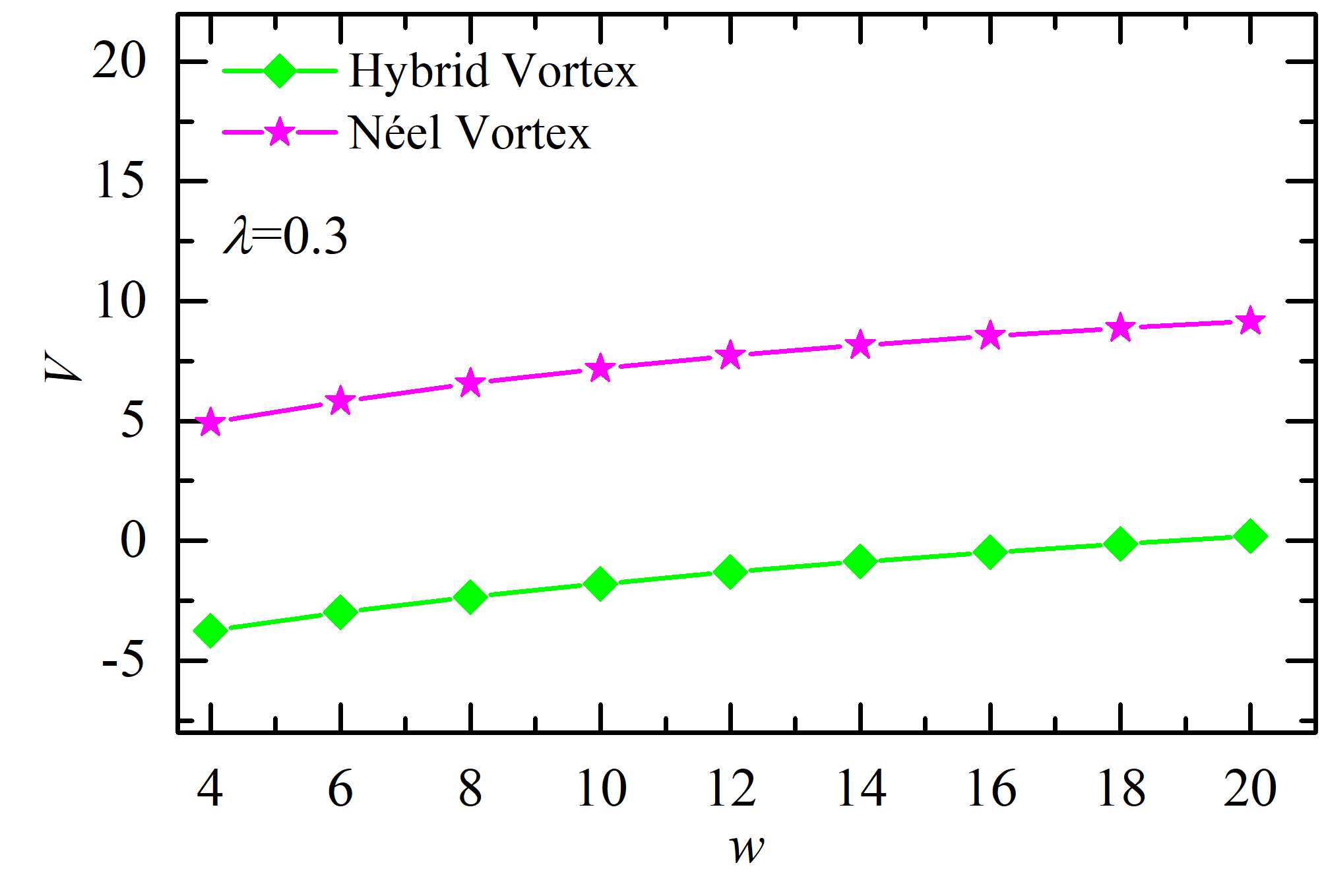}
    \caption{Energy $\Potential$ above the ground state energy as a function of the stripe width $\width$ for the hybrid vortex and for the N\'eel vortex for $\dm=0.3$.
    The numerical results are given by symbols (rhombus, star) connected by solid lines.
    }
    \label{fig:Energy_width}
\end{figure}

The vortex energy is finite in a stripe, in contrast to the logarithmically diverging vortex energy in infinite films.
The vortex energy above the ground state as a function of the stripe width $\width$ is shown in Fig.~\ref{fig:Energy_width}.
We find numerically a N\'eel-type vortex in the same stripe geometries by starting our relaxation simulations with a N\'eel vortex as an initial state.
Its energy, shown in Fig.~\ref{fig:Energy_width}, is higher than the energy of the hybrid vortex for the whole range of stripe widths $\width$.

\section{Propagating vortex}
\label{sec:propagatingVortex}

%\paragraph*{Boosted 1D state.}
We proceed to study the dynamics of the hybrid vortex.
Let us assume that a magnetic configuration is set into motion and we obtain $\nagn(x-\vel t)$ propagating along the axis of the stripe.
We initially neglect the dependence on $y$.
Eq.~\eqref{eq:sigmaModel} reduces to
\begin{equation} \label{eq:propagating1D}
    \nagn\times \left[(1-\vel^2) \p_1^2\nagn - 2\dm\ey\times \p_1\nagn - n_3 \ez \right] = 0.
\end{equation}
The latter equation is discussed in Appendix~\ref{sec:propagatingDW} in connection with propagating domain walls.
Applying a rescaling $x\to x \sqrt{1-\vel^2}$, Eq.~\eqref{eq:propagating1D} takes the form
\begin{equation} \label{eq:propagating1D_resc}
    \nagn\times \left( \p_1^2\nagn - 2\frac{\dm}{\sqrt{1-\vel^2}}\ey\times \p_1\nagn - n_3 \ez \right) = 0,
\end{equation}
where a single combination of parameters appears.
The phases of this 1D system were explained earlier in the introduction.
We have the following three cases.
(a) The N\'eel state for
\begin{equation} \label{eq:vNF}
\frac{\dm}{\sqrt{1-\vel^2}} < \dm_{NF} \Rightarrow \vel < \sqrt{1-\left(\frac{\dm}{\dm_{NF}}\right)^2}
 \equiv \vel_{NF}.
\end{equation}
(b) The non-flat spiral for
\begin{align}
& \dm_{NF} < \frac{\dm}{\sqrt{1-\vel^2}} < \dm_F \Rightarrow \notag \\
& \sqrt{1-\left(\frac{\dm}{\dm_{NF}}\right)^2} < \vel < \sqrt{1-\left(\frac{\dm}{\dm_F}\right)^2}.
\end{align}
(c) The flat spiral for
\begin{equation} \label{eq:vF}
\frac{\dm}{\sqrt{1-\vel^2}} > \dm_F \Rightarrow \vel > \sqrt{1-\left(\frac{\dm}{\dm_F}\right)^2} \equiv \vel_F.
\end{equation}

\begin{figure}
    \centering
    \includegraphics[width=8.5cm]{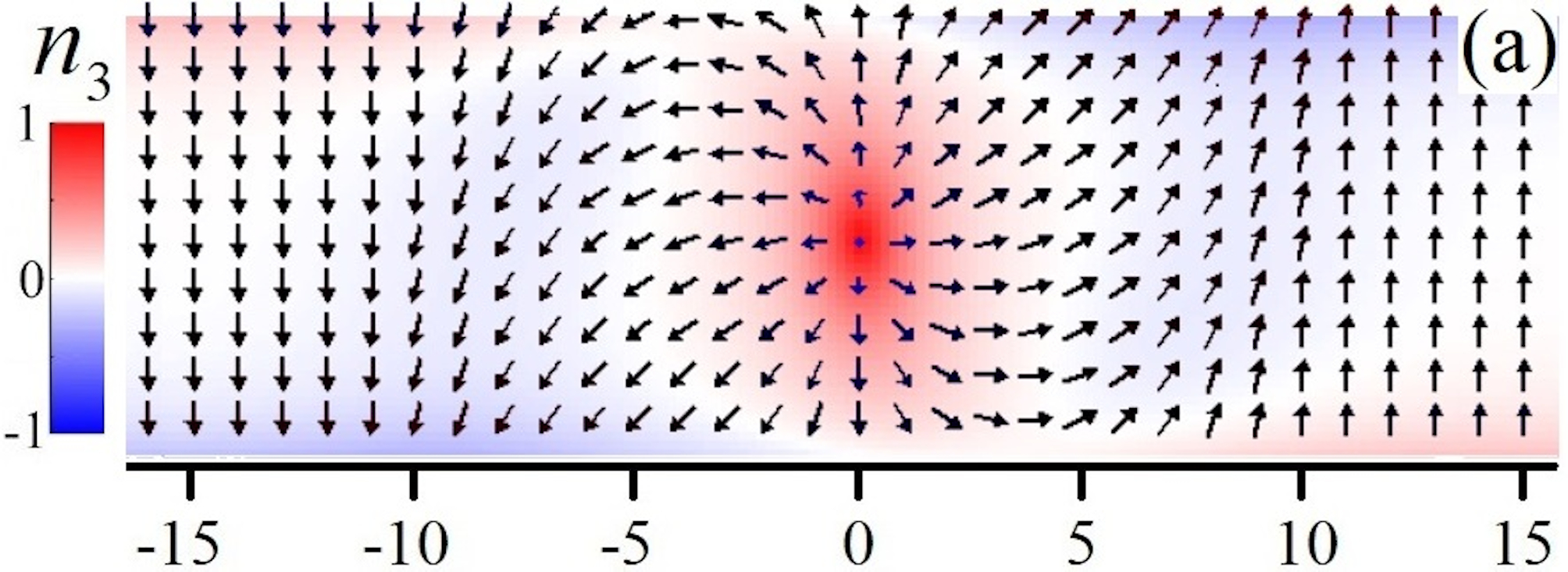}
    \vspace{15pt}
    
    \includegraphics[width=8.5cm]{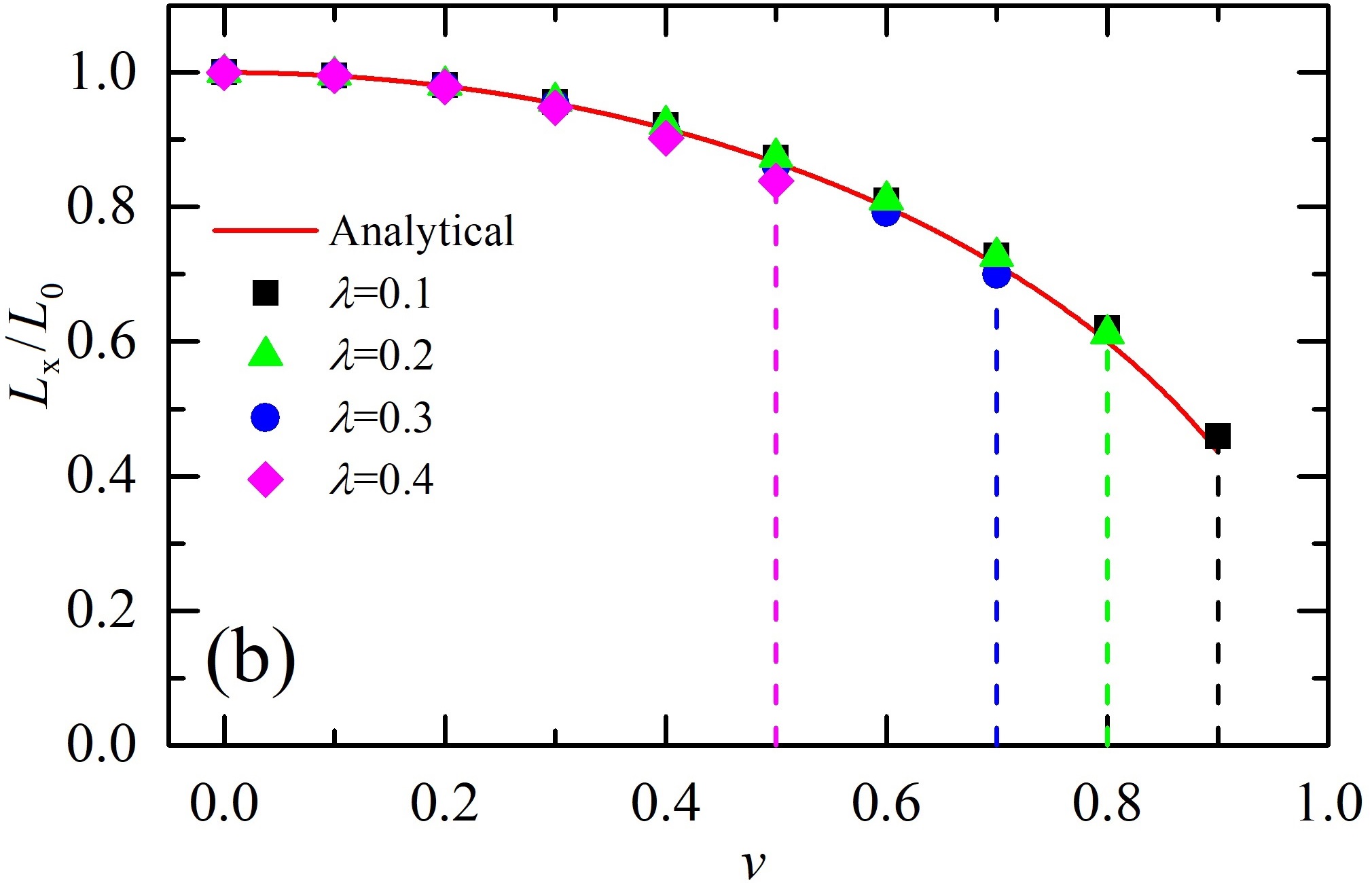}
    \vspace{15pt}
    
    \includegraphics[width=8.5cm]{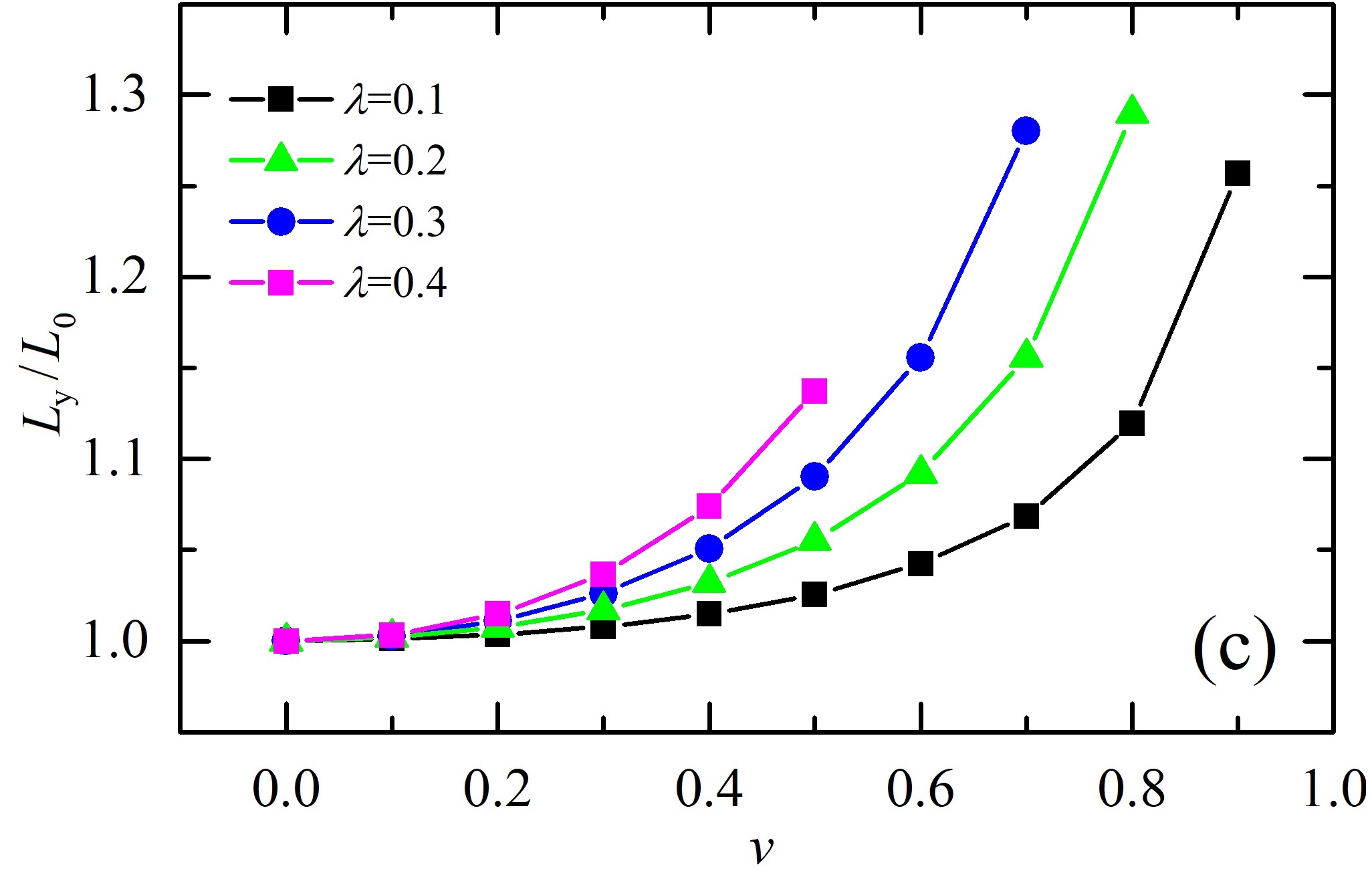}
    \caption{(a) Vector plot for $\dm=0.4$ for the propagating hybrid vortex for velocity $\vel=0.60$.
    Plotting conventions are as in Fig.~\ref{fig:vortexProfiles}.
    (b) Vortex core width $L_x$ in the direction of propagation for a propagating vortex as a function of velocity $\vel$, for various values of $\dm$, normalized to the width of a static vortex $L_0$.
    The red solid line shows the expected result for Lorentz-type contraction $L_x=\sqrt{1-\vel^2}$.
    The dashed lines mark the maximum obtained velocities for the respective $\dm$ values.
    (c) Vortex core width $L_y$ in the $y$ axis for the propagating vortex as a function of velocity $\vel$, normalized to the width of a static vortex $L_0$.
    }
    \label{fig:propagatingVortex}
\end{figure}

Using a numerical relaxation method  \cite{KomineasPapanicolaou_SciPost2020} applied to Eq.~\eqref{eq:propagating1D}, we find hybrid vortices in a steady-state motion propagating along the axis of the stripe with a range of velocities $\vel$.
Figure~\ref{fig:propagatingVortex}(a) shows a propagating vortex with velocity $\vel=0.6$.
Starting from the static hybrid vortex and increasing $\vel$, we find that the propagating vortex is contracted along the $x$ direction and it is elongated along the $y$ direction.

\begin{figure*}[t]
    \centering
    \includegraphics[width=17.5cm]{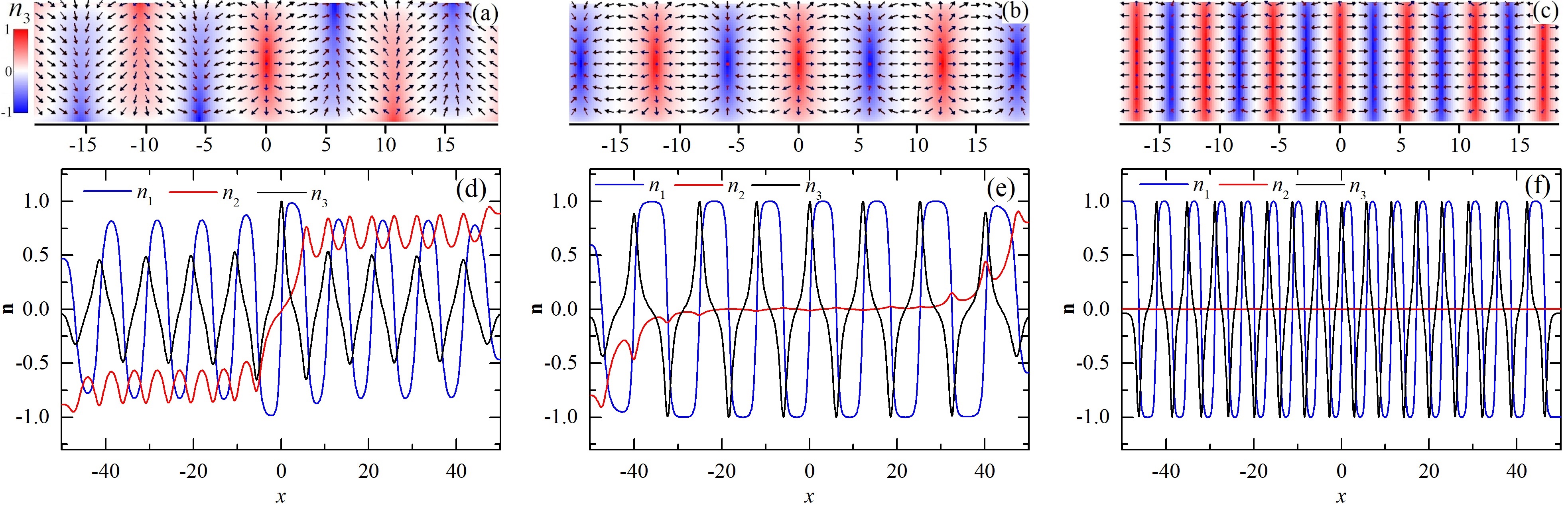}
    \caption{Vector plots for parameter $\dm=0.4$ for (a) a non-flat spiral at velocity $\vel=0.78$, (b) a vortex chain at $\vel=0.79$ and (c) a flat spiral at $\vel=0.90$.
    Plotting conventions are as in Fig.~\ref{fig:vortexProfiles}.
    (d)-(f) The components of $\nagn$ along the line in the center of the stripe ($y=0$) for the configurations shown in entries (a)-(c).
    The effect of the numerical mesh boundary, seen at $x=\pm50$ in entries (f,e), has a negligible effect on the configurations in the lattice interior.
    }
    \label{fig:periodicStates}
\end{figure*}

Figure~\ref{fig:propagatingVortex}(b) shows the width $L_x$ of the propagating vortex in the $x$ axis as a function of velocity for various values of $\dm$, normalized to the width $L_0$ of the static vortex.
We define the width of the vortex core as the distance between the positions where $n_3=0.5$.
The width $L_x$, in the direction of propagation, closely follows the law of Lorentz-type contraction (shown by a solid line in the figure) despite that the model is not Lorentz invariant.
Lorentz contraction is exactly followed by a propagating DW as reported in Ref.~\cite{Shiino2016} and reviewed in Appendix~\ref{sec:propagatingDW}.
For each $\dm$, the vortex achieves a maximum velocity (marked by dashed lines) as we explain below.
Therefore, there is a minimum achievable vortex width which decreases with decreasing $\dm$.
Figure~\ref{fig:propagatingVortex}(c) shows the width $L_y$ of the vortex core in the $y$ direction.
It increases with the velocity further pronouncing the vortex elongation.

When the velocity exceeds the value $\vel_{NF}(\dm)$ in Eq.~\eqref{eq:vNF}, we expect a nonflat spiral to develop based on the reasoning given following Eq.~\eqref{eq:propagating1D_resc}.
The numerical simulations show that this actually happens in the case of the stripe at a higher velocity.
Figure~\ref{fig:periodicStates}(a) shows the nonflat spiral that is nucleated, for $\dm=0.4$, when a single vortex is set into motion with a velocity $\vel=0.78$.
The vortex has survived in the stripe center and it is strongly elongated in the $y$ direction.
Figure~\ref{fig:periodicStates}(d) shows line plots of the N\'eel vector components along the line $y=0$ in the center of the stripe.
The spiral configuration is obvious in the $n_1, n_3$ components.
The component $n_2$ oscillates around nonzero values with opposite signs on the two sides of the vortex.
The configuration has the features of a DW on top of a spiral state (or a defect in the periodic structure).
Such a DW is connecting two topologically distinct spatially modulated ground states and it has been reported in Ref.~\cite{2004_APP_ChovanMarderPapanicolaou}.
Apart from the presence of a vortex in the center of the stripe, the structure is different from the ideal 1D nonflat spiral in that (a) $\nagn$ tilts out-of-plane close to the stripe boundaries and
(b) the spiral structure is different close to the stripe boundaries than in the stripe center as seen in the vector plot.
Indeed, edge (half) vortices are present at the boundaries of the stripe.

Further increasing the velocity, for large enough $\dm$, we obtain a periodic chain of vortices with opposite polarities, as shown in Fig.~\ref{fig:periodicStates}(b).
It appears that the edge vortices already present in Fig.~\ref{fig:periodicStates}(a) enter the stripe and develop into full vortices in Fig.~\ref{fig:periodicStates}(b).
The transition from the nonflat spiral to the chain of vortices appears to be a discontinuous one.
For example, in the transition between Figs.~\ref{fig:periodicStates}(a),(b), one can see the sudden change of the periodicity of the structure.
We have a phase transition to a lattice of topological solitons induced by the dynamics.

\begin{figure}
    \centering
    \includegraphics[width=8cm]{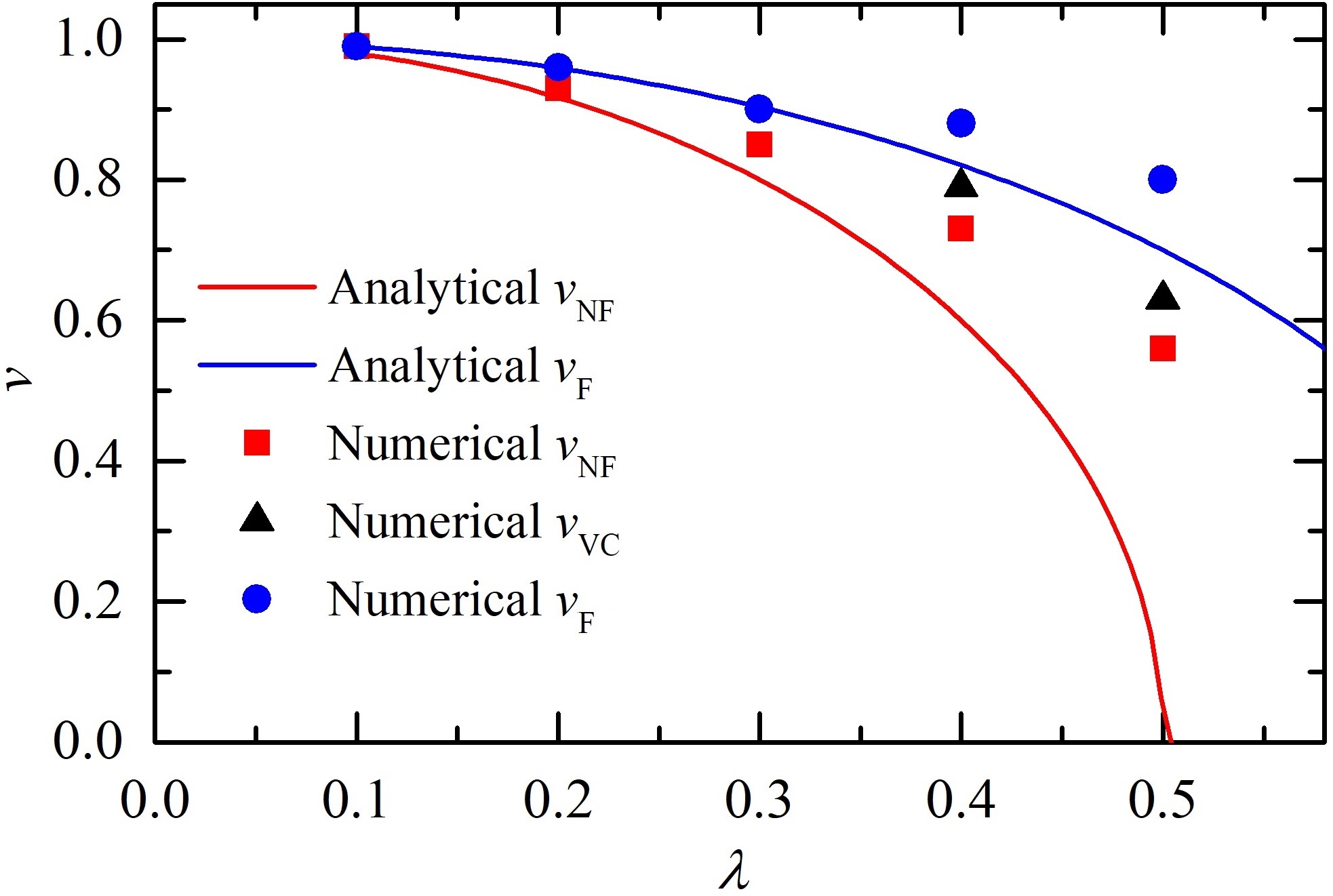}
    \caption{
    (a) The dots mark the numerically found velocities for the transition to the nonflat spiral (red squares), to the vortex chain (black triangles), and to the flat spiral (blue circles) for various values of $\dm$.
    The red solid line shows the velocity $\vel_{NF}(\dm)$ of Eq.~\eqref{eq:vNF} and the blue solid line shows $\vel_F(\dm)$ of Eq.~\eqref{eq:vF}, for comparison with the numerical results.
    }
    \label{fig:Crit_Vel}
\end{figure}

When the velocity exceeds the value $\vel_F(\dm)$ in Eq.~\eqref{eq:vF}, we expect a flat spiral to develop.
This actually happens close to $\vel=\vel_F(\dm)$ for small $\dm$ and for $\vel$ larger than $\vel_F(\dm)$ for large $\dm$.
Figure~\ref{fig:periodicStates}(c) shows a flat spiral in the stripe.
The vortex gets elongated across the width of the stripe and disappears from the configuration, while the component $n_2$ is nearly zero.
As a results, the configuration is close to the 1D spiral but some dependence of $\nagn$ on $y$ is seen in the region close to the boundaries. 
The transition from the chain of vortices to the spiral appears to be a continuous one.
Figure~\ref{fig:Crit_Vel} shows the numerically found velocities for the transitions to the nonflat spiral, the vortex chain and the flat spiral for various values of the DMI parameter $\dm$.
The velocities $\vel_{NF}(\dm), \vel_F(\dm)$ are plotted by solid lines for comparison.

Regarding the transition to the nonflat spiral, we attribute the deviations from the expected transition velocity to the 2D nature of the structure explained in connection with Fig.~\ref{fig:periodicStates}(a).
In a more quantitative argument, the boundary conditions favor the orientation of $\nagn$ in the $\ey$ over the $\ex$ direction, and it is therefore expected that the N\'eel state will persist longer, compared to the 1D model, before it is destabilized to the nonflat spiral.
Regarding the transition to the flat spiral, this is happening at $\vel$ larger than $\vel_F$ clearly due to the appearance of an additional state, that is, the vortex chain.
For small $\dm$, no vortex chain is formed because the transition to the nonflat spiral occurs at a high velocity $\vel\approx\vel_{\rm NF}$ where the vortex is very elongated.

\section{Concluding remarks}
\label{sec:conclusions}

We have studied vortices and their dynamics in an antiferromagnet with easy-plane anisotropy and interfacial DMI.
We have considered a nanostripe geometry and applied a continuum model.
The stripe boundary induces a quasi-uniform ground state with the N\'eel vector lying primarily perpendicular to the boundary.
The form of the ground state forces the vortex to have a hybrid character with both N\'eel and Bloch chirality.
When propagating, the hybrid vortex gives rise to phase transitions to a non-flat spiral, a vortex chain, and a flat spiral successively as the velocity increases.
While the spiral phases are anticipated by a study of the 1D model, the vortex chain is a feature of the stripe geometry.
No vortex lattice has been found in this system in an infinite film \cite{2002_PRB_ChovanPapanicolaou}.

\bigskip
\bigskip
\bigskip
\bigskip

\section*{Acknowledgements}
This work was supported by the project “ThunderSKY” funded by the Hellenic Foundation for Research and Innovation and the General Secretariat for Research and Technology, under Grant No. 871.
We acknowledge discussions with Michael Plexousakis on the numerical algorithms.

\newpage

\appendix

\section{One-dimensional system with boundaries}
\label{sec:groundState}

We assume a one-dimensional system of length $\width$, specifically, we consider a time-independent N\'eel vector $\nagn=\nagn(y)$ in the interval $-\width/2 \leq y \leq \width/2$.
This satisfies a reduced form of Eq. (2) of the main text,
\begin{equation} \label{eq:sigmaModel-y}
    \nagn\times (\nagn'' + 2\dm\ex\times\nagn' - n_3\ez) = 0
\end{equation}
where the prime denotes differentiation with respect to $y$.
The equation is supplemented with the boundary condition
\begin{equation} \label{eq:bc_1D-1}
    \nagn' + \dm \ex\times\nagn = 0,\quad
    y=\pm \frac{\width}{2}.
\end{equation}
We are looking for the ground state of this system.

An obvious solution of Eq.~\eqref{eq:sigmaModel-y} is $\nagn=\ex$ and this also satisfies the boundary condition.
Its energy is $\Energy=0$.

A state with negative energy can be found if we write
\begin{equation} \label{eq:flatConfiguration}
n_1 = 0,\quad
n_2 = \cos\Theta,\quad
n_3 = \sin\Theta
\end{equation}
where we use the parametrization with the polar angle $\Theta$ measured from the $\ey$ direction.
Eq. (1) of the main text for the energy reduces to the form
\begin{equation} \label{eq:potential_iDMI_ThetaPhi}
\Potential = \half \int (\Theta')^2\,\dy + \half \int \sin^2\Theta\,\dy + \dm \int \Theta'\,\dy
\end{equation}
where the integrations extend over the interval $-\width/2 \leq y \leq \width/2$.

Energy minimization, $\delta\Energy/\delta\Theta = 0$, gives
\begin{equation} \label{eq:boundary_equation0}
(\Theta')^2 = \sin^2\Theta + \gamma^2
\end{equation}
where $\gamma$ is a constant.
The boundary condition is
\begin{equation} \label{eq:bc_iDMI_Theta}
\frac{\delta\Potential}{\delta\Theta'} = 0
\Rightarrow \Theta' = -\dm,\qquad y=\pm \frac{\width}{2}
\end{equation}
and coincides with \eqref{eq:bc_1D-1}.
In the present problem, we will assume $\nagn(y=0) = \pm \ey$ in the center of the interval (the solution will be symmetric with respect to the center).
Thus, we confine the problem in the interval $0 \leq y \leq \width/2$ and we are seeking solutions with the boundary conditions
\begin{equation} \label{eq:bc_pi2}
\Theta(y=0) = 0, \pi \qquad
\Theta'\left(y=\pm\textstyle{\frac{\width}{2}}\right) = -\dm.
\end{equation}

\begin{figure}[t]
    \centering
    \includegraphics[width=8cm]{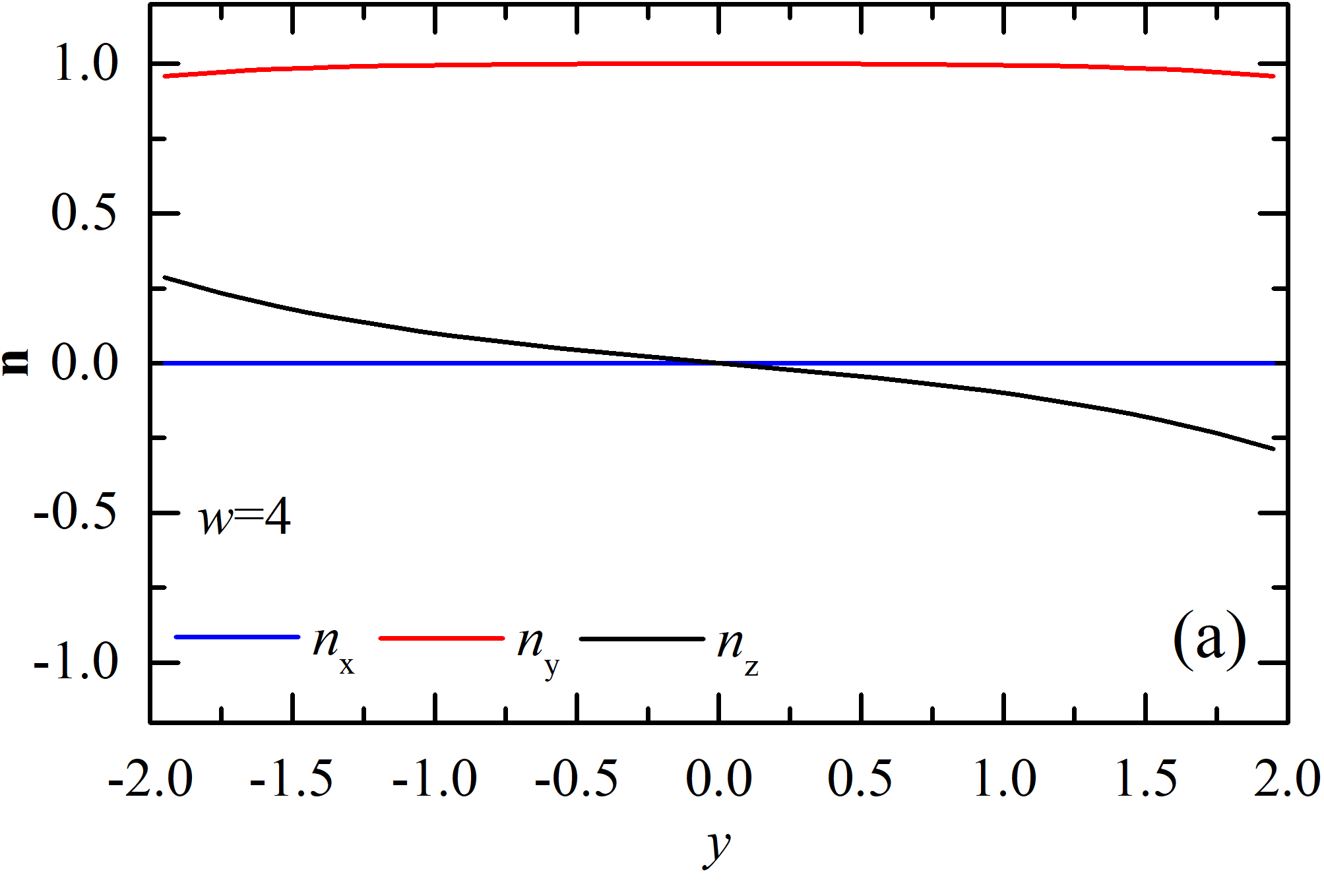}
    \bigskip
    
    \includegraphics[width=8cm]{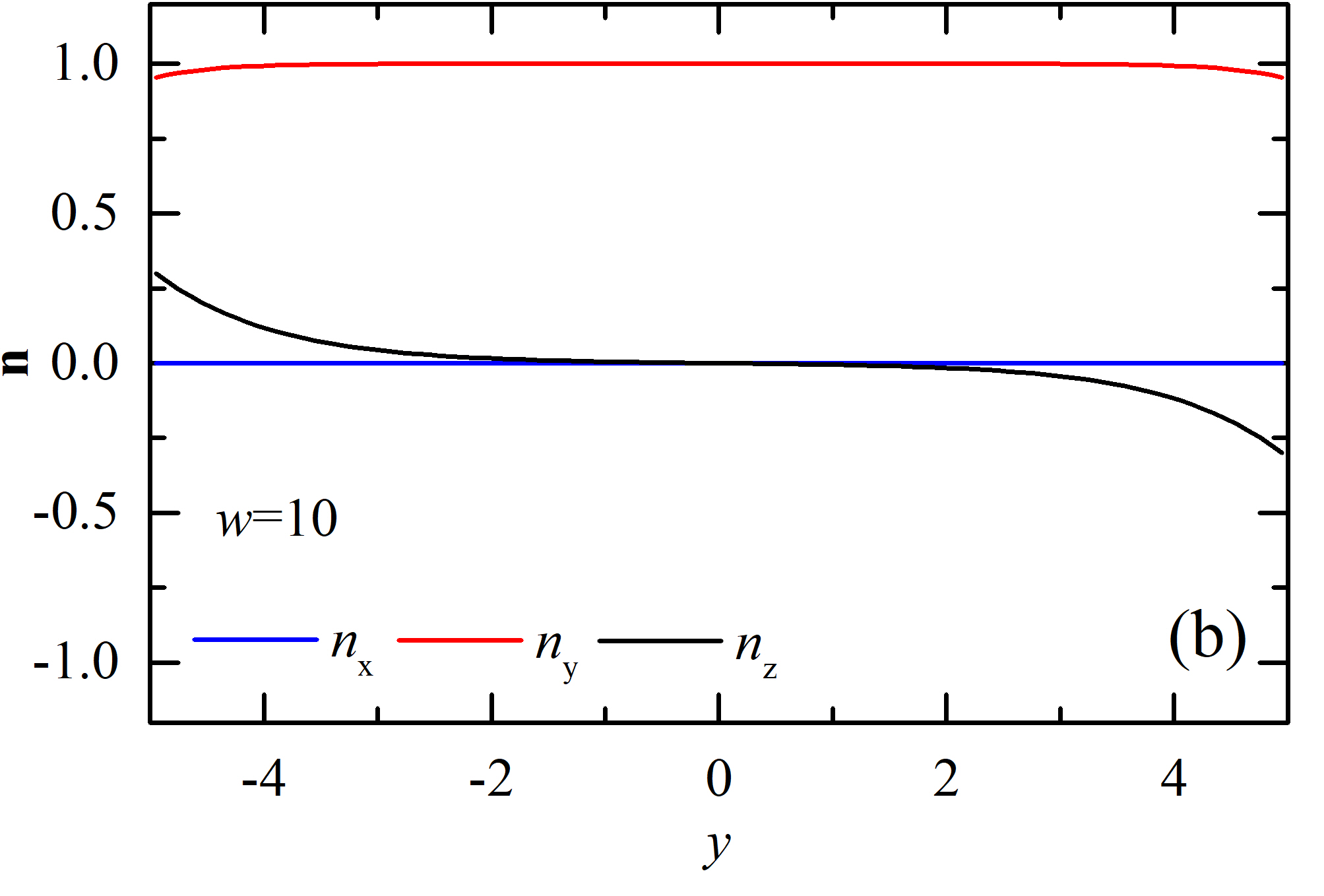}
    \caption{The ground state obtained as the solution of Eq.~\eqref{eq:sigmaModel-y} for the boundary conditions in Eq.~\eqref{eq:bc_1D-1} for system lengths $\width=4$ and 10.
    A second solution is obtained by $\nagn\to -\nagn$.
    We denote these two states by $\nagn_\pm$.}
    \label{fig:config1D_w4_w10}
\end{figure}

For the case $\Theta(y=0)=0$, Eq.~\eqref{eq:boundary_equation0} has the implicit solution
\begin{equation} \label{eq:formalSolution}
y = -\int_0^\Theta \frac{d\theta}{\sqrt{\sin^2\theta + \gamma^2}}
\end{equation}
where we have chosen the case $\Theta' < 0$ and thus $\Theta(y)$ is a monotonically decreasing function of $y$.
Fig.~\ref{fig:config1D_w4_w10} shows the components of $\nagn$ found numerically by solving Eq.~\eqref{eq:sigmaModel-y}, for two values of the system length $\width$.
We have a tilting of the N\'eel vector out-of-plane near the edges of the system.
The system is symmetric with respect to the transformation $\nagn\to -\nagn$.
The two equivalent solutions will be denoted by $\nagn_\pm$.

A remark of significant practical importance regarding the numerical application of the boundary conditions is the following.
We have found the solutions of \eqref{eq:sigmaModel-y} by using the boundary conditions \eqref{eq:bc_1D-1} and also by using open boundary conditions inspired by the physical problem.
In the latter case, the edge spins have only one neighbor.
The result for the states $\nagn_\pm$ is the same in both cases indicating that the two boundary conditions are equivalent (as shown in Appendix~\ref{sec:bc}).
This could be anticipated as the natural boundary conditions are indeed derived in order to describe free edges of the material.

We further denote
\begin{equation}
\Theta'(y=0) = \gamma,\qquad
\Theta\left(y=\textstyle{\pm\frac{\width}{2}}\right) = \mp\Theta_\width, \pi \mp\Theta_\width.
\end{equation}
At the boundaries, $y=\pm \width/2$, Eq.~\eqref{eq:boundary_equation0} gives the tilting angle $\Theta_\width$,
\begin{equation}
\sin^2\Theta_\width = \dm^2 - \gamma^2.
\end{equation}
This also implies that $|\gamma| < \dm$.

In the case of a {\it narrow stripe,} the angle is $|\Theta| \ll 1$ for all $y$ (assume the case $\Theta(y=0)=0$).
Eq.~\eqref{eq:formalSolution} gives
\[
\Theta(y) =  -\gamma y + O(y^3).
\]
To the same order of approximation, we have $\gamma \approx \dm$ and
\begin{equation}
\Theta(y) \approx -\dm y.
\end{equation}
The maximum angle, attained at the boundary, is $\Theta_\width = \dm\width/2$.
The condition for the validity of the result is $\dm\width \ll \gamma \Rightarrow \width \ll 1$.
%The corresponding result for the case $\Theta(y=0)=\pi$ is $\Theta(y)=\pi - \dm y$.
The energy \eqref{eq:potential_iDMI_ThetaPhi} has the value
\begin{equation} \label{eq:energy_smallew}
V = -\frac{\dm^2}{2} \width,\qquad \width \ll 1.
\end{equation}

In the case of a {\it wide stripe,} we assume that the configuration is almost uniform in the center, $\sin\Theta=0,\; \Theta'=0$.
We set $\gamma=0$ in Eq.~\eqref{eq:boundary_equation0} and this reduces to
\begin{equation} \label{eq:DW_equation}
(\Theta')^2 = \sin^2\Theta.
\end{equation}
Eq.~\eqref{eq:DW_equation} has the domain wall solution
\begin{equation} \label{eq:domainWall_Theta}
\tan\frac{\Theta}{2} =  -e^{y-y_0} 
%,\qquad\tan\frac{\Theta}{2} = e^{-(y-y_0)}
\end{equation}
where $y_0$ is a constant.
The N\'eel vector components are
\begin{equation} \label{eq:domainWall_n}
 n_2 = -\tanh(y-y_0), \quad n_3 = -\sech(y-y_0). %\\
%& n_2 = \cos\Theta  = \tanh(y-y_0),\quad\quad n_3 = \sin\Theta = \sech(y-y_0).
\end{equation}
The constant $y_0$ is determined by the boundary conditions \eqref{eq:bc_iDMI_Theta},
\begin{equation}
\Theta'(y=\pm \textstyle{\frac{\width}{2}}) = -\dm \Rightarrow
\sech \left(\pm\textstyle{\frac{\width}{2}}-y_0 \right) = \dm.
\end{equation}
At the boundaries, $\Theta'(\pm \width/2) = -\dm < 1/2$, thus $|y_0|>\width/2$ (that is, the center of the domain wall solution is beyond the boundary).
The form \eqref{eq:domainWall_n} applies to Fig.~\ref{fig:config1D_w4_w10} for $\width=10$.

\begin{figure}[t]
    \centering
    \includegraphics[width=8cm]{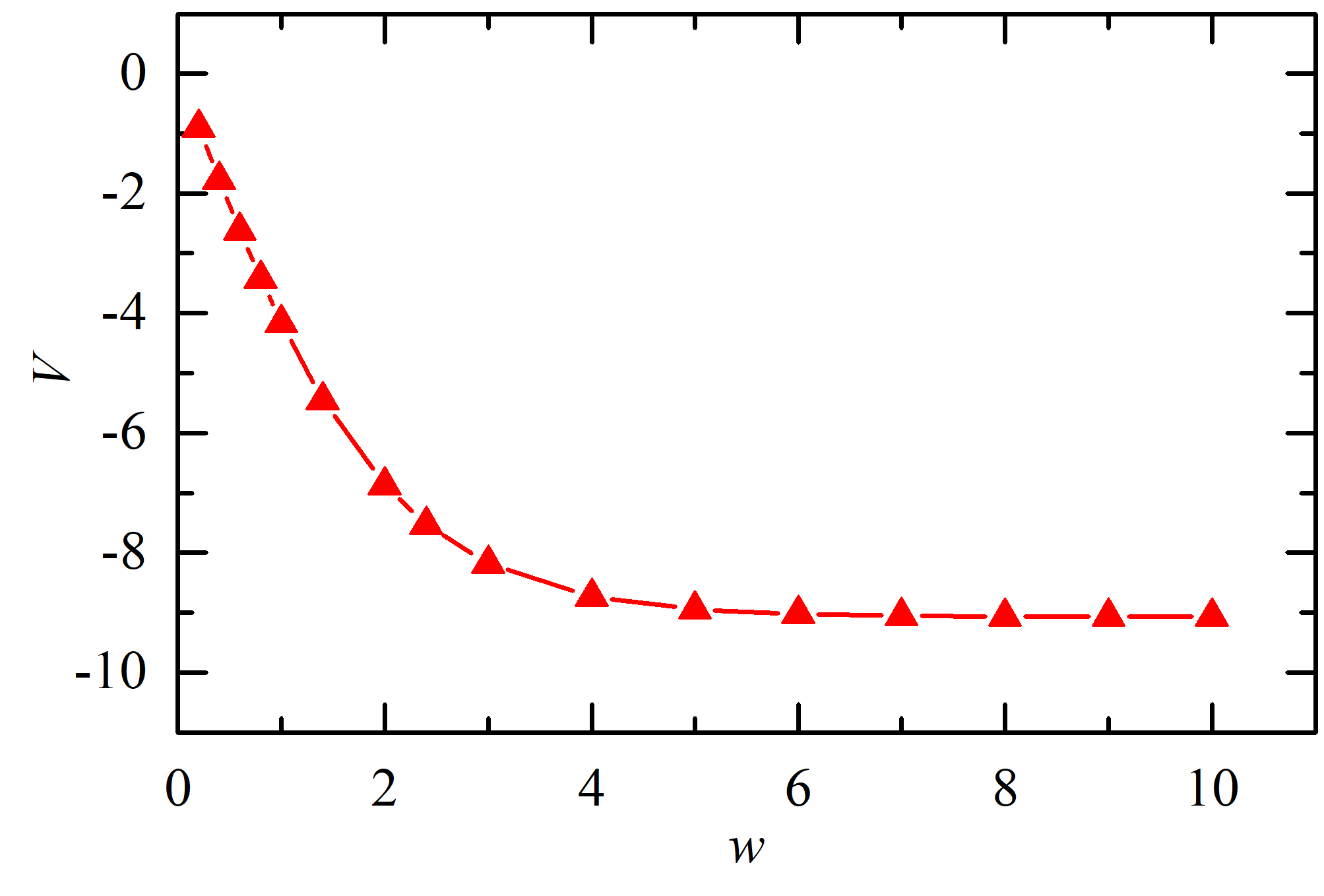}
    \caption{Energy $\Potential$ of the ground states $\nagn_\pm$ as a function of the system length $\width$ for $\dm=0.3$.
    Symbols show numerical results connected by a solid line.
    The slope of the curve for small $\width$ is given by Eq.~\eqref{eq:energy_smallew}.
    }
    \label{fig:E_w_1D}
\end{figure}

We will now prove that the energy for all $\nagn_\pm$ is $\Potential < 0$.
For $0 \leq \Theta \leq \pi/2$, Eq.~\eqref{eq:boundary_equation0} gives that $|\Theta'|$ is an increasing function of $\Theta$ and thus also an increasing function of $y$.
We have $|\Theta'(y)| \leq \dm$ with the maximum value attained at the boundary, $|\Theta'(y=\width/2)|=\dm$.
We insert Eq.~\eqref{eq:boundary_equation0} in Eq.~\eqref{eq:potential_iDMI_ThetaPhi} and then use the inequality for $|\Theta'|$ to find that the energy of the configuration is negative,
\begin{equation} \label{eq:energyNegative_1D}
\Potential \leq \int (\Theta')^2\,\dy + \dm \int \Theta'\,\dy < 0,
\end{equation}
where we take into account that $\Theta' < 0$.
Eq.~\eqref{eq:energyNegative_1D} establishes that the nonuniform states $\nagn_\pm$ have an energy lower than any uniform state in the system.
They are found numerically to be the lowest energy states.

Fig.~\ref{fig:E_w_1D} shows the energy of the ground states $\nagn_\pm$ as a function of the system length $\width$.
The dependence is linear for small $\width$, following Eq.~\eqref{eq:energy_smallew}, and it saturates to a negative value for larger $\width$.

\section{Propagating domain wall}
\label{sec:propagatingDW}

Let us consider the 1D system that results from Eq. (2) of the main text when we assume $\nagn=\nagn(x,t)$,
\begin{equation}
    \nagn\times \left( \ddot{\nagn} - \nagn'' + 2\dm \ey\times\nagn' + n_3 \ez \right)
\end{equation}
where the prime denotes differentiation with respect to $x$.
Denote by
\[
\nDW(x) = (\sech(x),0,\tanh(x))
\]
the static domain wall solution.
This is stable for $\dm<\dm_{NF}=\frac{1}{2}$ while for $\dm>\dm_{NF}$ it is destabilized to the nonflat spiral.

A domain wall propagating with velocity $\vel$ satisfies
\begin{equation} \label{eq:propagating1D-1}
    \nagn\times \left[ (1-\vel^2) \nagn''- 2\dm \ey\times\nagn' - n_3 \ez \right] = 0.
\end{equation}
The solution of the equation is obtained by a Lorentz transformation of the static wall
\[
\nagn(x,t;\vel)=\nDW\left(\frac{x-\vel t}{\sqrt{1-\vel^2}} \right).
\]
Note that the DM term vanishes for the static or propagating domain wall solutions and thus Lorentz invariance is preserved.
The propagating solution is valid for the range of parameter values where the N\'eel state is stable,
\begin{equation} \label{eq:vel0}
\frac{\dm}{\sqrt{1-\vel^2}} < \dm_{NF} \Rightarrow \vel < \sqrt{1-\left( \frac{\dm}{\dm_{NF}} \right)^2} \equiv \vel_0.
\end{equation}
As the velocity increases, the domain wall is contracted by a factor $\sqrt{1-\vel^2}$ and it has a minimum width at $\vel=\vel_0$. 
For $\vel > \vel_0$ the propagating domain wall is unstable and the system should turn to a propagating spiral state.

\section{Open and natural boundary conditions}
\label{sec:bc}

We consider a vector field $\nagn=\nagn(x,t)$ with components $\nagn=(n_1,n_2,n_3)$ and a constant length $|\nagn|=1$.
It satisfies the equation
\begin{equation} \label{eq:sigmaModel-1}
    \nagn\times \left( \nagn'' + 2\dm \ey\times\nagn' - n_3\ez \right) = 0
\end{equation}
where $\dm$ is a parameter.
The problem is defined in an interval $-\width/2 \leq x \leq \width/2$ and the boundary conditions (so-called, {\it natural boundary conditions}) are
\begin{equation} \label{eq:bc_1D-2}
    \nagn' + \dm \ey\times\nagn = 0,\qquad x=\pm\frac{\width}{2}.
\end{equation}

We discretise space and have a lattice of points $x_i,\, i=1,\ldots,N$ with lattice spacing $a$.
On the lattice, the discrete version of Eq.~\eqref{eq:sigmaModel-1} reads
\begin{equation} \label{eq:sigma_discrete-1}
    \nagn_i\times \left( \frac{\nagn_{i+1}+\nagn_{i-1}}{a^2} + \dm \ey\times \frac{\nagn_{i+1}-\nagn_{i-1}}{a} - n_{i,3}\ez \right) = 0
\end{equation}
for any site $i=1,\ldots,N$ of the lattice with lattice spacing $a$.

We consider the following two approaches for implementing the boundary conditions.

\paragraph{Open boundary conditions.}
Motivated by the physical problem, we use open boundary conditions, that is, we assume that there is no interaction to the right of the last site $i=N$, and thus Eq.~\eqref{eq:sigma_discrete-1} gives at the last site, $i=N$,
\begin{equation} \label{eq:signa_discrete_boundary-1}
    \nagn_N\times \left( \frac{\nagn_{N-1}}{a^2} - \dm \ey\times \frac{\nagn_{N-1}}{a} - n_{N,3}\ez \right) = 0.
\end{equation}
A similar equation is obtained for the first site $i=1$.

\paragraph{Apply bc's to order $O(a)$.}
The discrete form of \eqref{eq:bc_1D-2} at $i=N$ reads
\begin{align} \label{eq:bc_idm_a}
    & \frac{\nagn_{N+1}-\nagn_N}{a} + \dm\ey\times\nagn_{N+1} = 0  \notag \\
    \Rightarrow & \frac{\nagn_{N+1}}{a} + \dm\ey\times\nagn_{N+1} = \frac{\nagn_N}{a},
\end{align}
correct to order $O(a)$.
The latter can be used in Eq.~\eqref{eq:sigma_discrete-1} to give \eqref{eq:signa_discrete_boundary-1}.
This proves the equivalence of the open boundary conditions with the natural boundary conditions.

\bigskip

%\bibliographystyle{prsty}
%\bibliographystyle{aipnum4-1}
%\bibliography{references}

\begin{thebibliography}{10}

\bibitem{Jungwirth2016}
T. Jungwirth, X. Marti, P. Wadley, and J. Wunderlich, Nature Nanotechnology
  {\bf 11},  231  (2016).

\bibitem{Baltz2018}
V. Baltz, A. Manchon, M. Tsoi, T. Moriyama, T. Ono, and Y. Tserkovnyak, Reviews
  of Modern Physics {\bf 90},  015005  (2018).

\bibitem{Wadley2016}
P. Wadley, B. Howells, J. Elezny, C. Andrews, V. Hills, R.~P. Campion, V.
  Novak, K. Olejnik, F. Maccherozzi, S.~S. Dhesi, S.~Y. Martin, T. Wagner, J.
  Wunderlich, F. Freimuth, Y. Mokrousov, J. Kune, J.~S. Chauhan, M.~J.
  Grzybowski, A.~W. Rushforth, K.~W. Edmonds, B.~L. Gallagher, and T.
  Jungwirth, Science {\bf 351},  587  (2016).

\bibitem{Grzybowski2017}
M.~J. Grzybowski, P. Wadley, K.~W. Edmonds, R. Beardsley, V. Hills, R.~P.
  Campion, B.~L. Gallagher, J.~S. Chauhan, V. Novak, T. Jungwirth, F.
  Maccherozzi, and S.~S. Dhesi, Physical Review Letters {\bf 118},  057701
  (2017).

\bibitem{Moriyama2018}
T. Moriyama, K. Oda, T. Ohkochi, M. Kimata, and T. Ono, Scientific Reports {\bf
  8},  14167  (2018).

\bibitem{Bodnar2019}
S.~Y. Bodnar, M. Filianina, S.~P. Bommanaboyena, T. Forrest, F. Maccherozzi,
  A.~A. Sapozhnik, Y. Skourski, M. Kl{\"{a}}ui, and M. Jourdan, Physical Review
  B {\bf 99},  140409  (2019).

\bibitem{Baldrati2019}
L. Baldrati, O. Gomonay, A. Ross, M. Filianina, R. Lebrun, R. Ramos, C.
  Leveille, F. Fuhrmann, T.~R. Forrest, F. Maccherozzi, S. Valencia, F.
  Kronast, E. Saitoh, J. Sinova, and M. Kl{\"{a}}ui, Physical Review Letters
  {\bf 123},  177201  (2019).

\bibitem{Shi2020}
J. Shi, V. Lopez-Dominguez, F. Garesci, C. Wang, H. Almasi, M. Grayson, G.
  Finocchio, and P. {Khalili Amiri}, Nature Electronics {\bf 3},  92  (2020).

\bibitem{Cheng2015}
R. Cheng, M.~W. Daniels, J.-G. Zhu, and D. Xiao, Physical Review B {\bf 91},
  064423  (2015).

\bibitem{Roy2016}
P.~E. Roy, R.~M. Otxoa, and J. Wunderlich, Physical Review B {\bf 94},  014439
  (2016).

\bibitem{Lopez-Dominguez2019}
V. Lopez-Dominguez, H. Almasi, and P.~K. Amiri, Physical Review Applied {\bf
  11},  024019  (2019).

\bibitem{Cheng2016}
R. Cheng, D. Xiao, and A. Brataas, Physical Review Letters {\bf 116},  207603
  (2016).

\bibitem{Khymyn2017}
R. Khymyn, I. Lisenkov, V. Tiberkevich, B.~A. Ivanov, and A. Slavin, Scientific
  Reports {\bf 7},  43705  (2017).

\bibitem{Puliafito2019}
V. Puliafito, R. Khymyn, M. Carpentieri, B. Azzerboni, V. Tiberkevich, A.
  Slavin, and G. Finocchio, Physical Review B {\bf 99},  024405  (2019).

\bibitem{Gomonay2016}
O. Gomonay, T. Jungwirth, and J. Sinova, Physical Review Letters {\bf 117},
  017202  (2016).

\bibitem{Shiino2016}
T. Shiino, S.-H. Oh, P.~M. Haney, S.-W. Lee, G. Go, B.-G. Park, and K.-J. Lee,
  Physical Review Letters {\bf 117},  087203  (2016).

\bibitem{Sanchez-Tejerina2020}
L. S{\'{a}}nchez-Tejerina, V. Puliafito, P. {Khalili Amiri}, M. Carpentieri,
  and G. Finocchio, Physical Review B {\bf 101},  014433  (2020).

\bibitem{Zhang2016}
X. Zhang, Y. Zhou, and M. Ezawa, Scientific Reports {\bf 6},  24795  (2016).

\bibitem{Barker2016}
J. Barker and O.~A. Tretiakov, Physical Review Letters {\bf 116},  147203
  (2016).

\bibitem{Gomonay2018}
O. Gomonay, V. Baltz, A. Brataas, and Y. Tserkovnyak, Nature Physics {\bf 14},
  213  (2018).

\bibitem{Salimath2020}
A. Salimath, F. Zhuo, R. Tomasello, G. Finocchio, and A. Manchon, Physical
  Review B {\bf 101},  024429  (2020).

\bibitem{IvanovSheka_PRL1994}
B.~A. Ivanov and D.~D. Sheka, Phys. Rev. Lett. {\bf 72},  404  (1994).

\bibitem{Pereira1995}
A.~R. Pereira and A.~S.~T. Pires, Physical Review B {\bf 51},  996  (1995).

\bibitem{Ivanov1996}
B.~A. Ivanov, A.~K. Kolezhuk, and G.~M. Wysin, Physical Review Letters {\bf
  76},  511  (1996).

\bibitem{Bogdanov1998}
A. Bogdanov and A. Shestakov, Physics of the Solid State {\bf 40},  1350
  (1998).

\bibitem{Komineas1998}
S. Komineas and N. Papanicolaou, Nonlinearity {\bf 11},  265  (1998).

\bibitem{Wu2011}
J. Wu, D. Carlton, J.~S. Park, Y. Meng, E. Arenholz, A. Doran, A.~T. Young, A.
  Scholl, C. Hwang, H.~W. Zhao, J. Bokor, and Z.~Q. Qiu, Nature Physics {\bf
  7},  303  (2011).

\bibitem{Chmiel2018}
F.~P. Chmiel, N. {Waterfield Price}, R.~D. Johnson, A.~D. Lamirand, J. Schad,
  G. van~der Laan, D.~T. Harris, J. Irwin, M.~S. Rzchowski, C.-B. Eom, and
  P.~G. Radaelli, Nature Materials {\bf 17},  581  (2018).

\bibitem{1989_PRB_GouveaWysinBishopMertens}
M.~E. Gouv\^ea, G.~M. Wysin, A.~R. Bishop, and F.~G. Mertens, Phys. Rev. B {\bf
  39},  11840  (1989).

\bibitem{PapanicolaouSpathis_NL1999}
N. Papanicolaou and P.~N. Spathis, Nonlinearity {\bf 12},  285  (1999).

\bibitem{2000_Science_Shinjo}
T. Shinjo, T. Okuno, R. Hassdorf, {\textdagger}.~K. Shigeto, and T. Ono,
  Science {\bf 289},  930  (2000).

\bibitem{2002_Science_Wachowiak}
A. Wachowiak, J. Wiebe, M. Bode, O. Pietzsch, M. Morgenstern, and R.
  Wiesendanger, Science {\bf 298},  577  (2002).

\bibitem{WaeyenbergePuzic_Nat2006}
B.~V. Waeyenberge, A. Puzic, H. Stoll, K.~W. Chou, T. Tyliszczak, R. Hertel, M.
  F\"ahnle, H. Br\"uckl, K. Rott, G. Reiss, I. Neudecker, D. Weiss, C.~H. Back,
  and G. Sch\"utz, Nature(London) {\bf 444},  461  (2006).

\bibitem{Yamada2007}
K. Yamada, S. Kasai, Y. Nakatani, K. Kobayashi, H. Kohno, A. Thiaville, and T.
  Ono, Nature materials {\bf 6},  269  (2007).

\bibitem{Pribiag2007}
V.~S. Pribiag, I.~N. Krivorotov, G.~D. Fuchs, P.~M. Braganca, O. Ozatay, J.~C.
  Sankey, D.~C. Ralph, and R.~A. Buhrman, Nature Physics {\bf 3},  498  (2007).

\bibitem{Komineas2007}
S. Komineas, Physical Review Letters {\bf 99},  117202  (2007).

\bibitem{2011_PRB_MuhlbauerZheludev}
S. M\"uhlbauer, S.~N. Gvasaliya, E. Pomjakushina, and A. Zheludev, Phys. Rev. B
  {\bf 84},  180406  (2011).

\bibitem{2012_PRB_MuhlbauerZheludev}
S. M\"uhlbauer, S. Gvasaliya, E. Ressouche, E. Pomjakushina, and A. Zheludev,
  Phys. Rev. B {\bf 86},  024417  (2012).

\bibitem{2002_PRB_ChovanPapanicolaou}
J. Chovan, N. Papanicolaou, and S. Komineas, Phys. Rev. B {\bf 65},  064433
  (2002).

\bibitem{2005_springer_ChovanPapanicolaou}
J. Chovan and N. Papanicolaou,  in {\em Frontiers in Magnetic Materials},
  edited by A.~V. Narlikar (Springer Berlin Heidelberg, Berlin, Heidelberg,
  2005), pp.\ 347--384.

\bibitem{BogdanovHubert_JMMM1999}
A.~N. Bogdanov and A. Hubert, JMMM {\bf 195},  182  (1999).

\bibitem{KomineasPapanicolaou_SciPost2020}
S. Komineas and N. Papanicolaou, SciPost Phys. {\bf 8},  086  (2020).

\bibitem{BaryakhtarIvanov_SJLTP1979}
I.~V. Baryakhtar and B.~A. Ivanov, Sov. J. of Low Temp. Phys. {\bf 5},  361
  (1979).

\bibitem{KomineasPapanicolaou_NL1998}
S. Komineas and N. Papanicolaou, Nonlinearity {\bf 11},  265  (1998).

\bibitem{Tomasello2020}
R. Tomasello, L. Sanchez-Tejerina, V. Lopez-Dominguez, F. Garesc{\`{i}}, A.
  Giordano, M. Carpentieri, P.~K. Amiri, and G. Finocchio, Physical Review B
  {\bf 102},  224432  (2020).

\bibitem{Thiaville2012}
A. Thiaville, S. Rohart, {\'{E}}. Ju{\'{e}}, V. Cros, and A. Fert, EPL
  (Europhysics Letters) {\bf 100},  57002  (2012).

\bibitem{Buttner2018}
F. B{\"{u}}ttner, I. Lemesh, and G.~S.~D. Beach, Scientific Reports {\bf 8},
  4464  (2018).

\bibitem{Olleros-Rodriguez2020}
P. Olleros-Rodr{\'{i}}guez, R. Guerrero, J. Camarero, O. Chubykalo-Fesenko, and
  P. Perna, ACS Applied Materials {\&} Interfaces {\bf 12},  25419  (2020).

\bibitem{2004_APP_ChovanMarderPapanicolaou}
J. Chovan, M. Marder, and N. Papanicolaou, Acta Physica Polonica {\bf 126},  32
   (2004).

\end{thebibliography}

\end{document}